\documentclass[pra,twocolumn,showpacs,floatfix]{revtex4}
\usepackage{graphicx}
\usepackage{amsmath}

\begin{document}

\title{Spin-orbit-coupled Bose-Einstein-condensed atoms confined in annular potentials}
\author{E. \"O. Karabulut$^1$, F. Malet$^2$, A. L. Fetter$^3$, G. M. Kavoulakis$^4$, 
and S. M. Reimann$^5$}
\affiliation{$^1$Physics Department, Faculty of Science, Selcuk University, TR-42075, 
Konya, Turkey \\
$^2$Department of Theoretical Chemistry and Amsterdam Center for Multiscale Modeling, 
FEW, Vrije Universiteit, De Boelelaan 1083, 1081HV Amsterdam, The Netherlands \\
$^3$Departments of Physics and Applied Physics, Stanford University, Stanford, CA 94305-4045, 
USA \\
$^4$Technological Educational Institute of Crete, P.O. Box 1939, GR-71004, Heraklion, Greece \\
$^5$Mathematical Physics, LTH, Lund University, P. O. Box 118, SE-22100 Lund, Sweden}

\date{\today}

\begin{abstract}

A spin-orbit-coupled Bose-Einstein-condensed cloud of atoms confined in an annular trapping 
potential shows a variety of phases that we investigate in the present study. Starting with
the non-interacting problem, the homogeneous phase that is present in an untrapped system is 
replaced by a sinusoidal density variation in the limit of a very narrow annulus. In the case
of an untrapped system there is another phase with a striped-like density distribution, and 
its counterpart is also found in the limit of a very narrow annulus. As the width of the annulus 
increases, this picture persists qualitatively. Depending on the relative strength between the 
inter- and the intra-components, interactions either favor the striped phase, or suppress it, 
in which case either a homogeneous, or a sinusoidal-like phase appears. Interactions also give 
rise to novel solutions with a nonzero circulation.

\end{abstract}


\maketitle

\section{Introduction}
 
The effects associated with spin-orbit coupling have long been known and studied in 
various physical systems, including atoms, solids, quantum dots, etc. \cite{dre, mir, kas, 
and, gov, wei, lip}, and nuclei \cite{nuclei}. Spin-orbit coupling often plays an important 
role in semiconductor nanostructures, where the field of spintronics has led to many new 
applications \cite{zut}. Concepts known from semiconductor nanoscience are nowadays often 
found to be transferable to confined atomic quantum gases, with an ever increasing interest 
in the properties of ``atomtronic" devices that make use of cold atoms in a similar way like 
electrons in solids \cite{at1, at2, at3, at4, at5, at6, at7, at8, at9}. Semiconductor 
``spintronic" devices are to a large extent controlled by the Rashba or Dresselhaus spin-orbit 
coupling in the heterostructure, giving rise to a number of intriguing spin-dependent transport 
phenomena \cite{spintr}. By means of laser-coupling techniques an analogue to spintronics has 
now become possible with the achievement of artificially-induced spin-orbit coupling in ultracold 
quantum gases of neutral atoms \cite{ost} with the exciting possibility of creating 
``atom-spintronic" devices that may also make use of the superfluid properties of quantum 
gases. Recently, such systems have been experimentally realized in Bose \cite{lin0, lin}, 
as well as in Fermi \cite{wang} gases. Their unique properties and their fundamental and 
application-related prospects have inspired intense theoretical efforts \cite{spth, merkl, wan, 
ho, xu, mxu2, rad, wu11, zhou11, zha, li, and2, gal, har2, li11, zha2, ram, and3, xu3, xu4, aft, 
den, zhou22, zez, che, kar11, lar, qu}. Furthermore, in a series of other experiments, it has become 
possible to trap atoms in annular/toroidal traps, see, e.g., Refs.\,\cite{gup, murray, wright}. 

Motivated by the above experiments, we investigate the lowest-energy states that appear in 
a spin-orbit-coupled Bose-Einstein-condensed cloud of atoms that is trapped in an annular 
potential, using the mean-field approximation. A lot of work has been done in homogeneous 
\cite{wan, ho, jia, zhai, oza, oza2} and in harmonically-trapped systems \cite{har2, xu, 
zha, zha2, ram, aft, sinha, huram, huliu}. In the presence of an annular potential, however, 
this problem becomes surprisingly challenging. As we show below, even in the absence of 
interactions the eigenvalue problem due to the spin-orbit coupling is non-trivial. 

We proceed in three steps: First of all, we ignore the interactions and consider the case 
of a very tight annulus, developing an effectively one-dimensional theory. Within this 
model we integrate over the transverse degrees of freedom, which not only simplifies the 
problem numerically, but it also gives some additional insight into the properties of the
system. Furthermore (Sec.\,III), the resulting equations have some limiting analytic 
solutions. We then solve the eigenvalue problem in an annulus of a finite width (Sec.\,IV). 
The interactions are incorporated in the last step (Sec.\,V). We pay special attention to 
the interplay between the (single-particle) effect of spin-orbit coupling and the atom-atom 
interactions; we stress that the parameters that are associated with both of them are realistic 
and controllable experimentally in atomic systems. 

\section{Model and method}

We consider a Bose-Einstein-condensed cloud of atoms, confined to two dimensions. Thus, the 
single-particle Hamiltonian of the system is
\begin{eqnarray}
H_0 = \sum_{i=1}^N \left[ - \hbar^2 \nabla_i^2/(2 M) + V(\vec{r}_i) + V_{SO}(\vec{r}_i) \right]. 
\label{hamilt}
\end{eqnarray}
The external confining potential has the form of an annular potential, $V(\vec{r}) = M \omega^2 
(\rho - R)^2/2$, where $\rho$ is the radial coordinate, $M$ is the atom mass, $\omega$ is 
the frequency of the potential, and $R$ is the mean radius of the annulus. The term $V_{SO}$ 
associated with the spin-orbit coupling that we consider here is the one that corresponds to the 
experiments of the Spielman group \cite{lin}. In this case the so-called Rashba-like and 
Dresselhaus-like spin-orbit contributions have an equal strength, and thus $V_{SO}$ takes the 
one-dimensional form 
\begin{eqnarray}
V_{SO}(\vec{r}) = \frac {\hbar^2 k_0^2} {2 M} + \frac {\hbar k_0 p_x} M \sigma^y + 
\frac 1 2 \hbar \Omega \sigma^z + \frac 1 2 \hbar \delta \sigma^y.
\label{soham}
\end{eqnarray} 
Here $k_0$ is the wavenumber of the Raman laser beams, $p_x$ is the linear momentum of the 
atoms in the $x$ direction, $\sigma^j$ (with $j=x,y,z$) are the Pauli matrices, $\Omega$ is 
the Rabi frequency and $\delta$ is the detuning. The experimental studies of spin-orbit 
coupling generally choose small values of $\delta$ \cite{lin0, lin} and thus we ignore the
corresponding term in what follows below. Furthermore, we here orient the spin axes as in 
Ref.\,\cite{lin}, but most recent theoretical work \cite{spth, zhai} uses a cyclic spin 
rotation $\sigma_y \to \sigma_z$, etc. Therefore, the system is described as an effective 
pseudo-spin-1/2, and the condensate wave function is written as a two-component vector 
$[\Phi_{\uparrow}, \Phi_{\downarrow}]^T$ (``up" and ``down"). 

The interactions are modelled via a short-ranged $s$-wave potential of the form 
\begin{eqnarray}
V_{\rm int} = \frac g 2 \sum_{s = \uparrow, \downarrow} \int |\Phi_{s}|^4 \, dx dy 
+ g_{\uparrow \downarrow} \int |\Phi_{\uparrow}|^2 |\Phi_{\downarrow}|^2 \, dx dy.
\label{interac} 
\end{eqnarray}
While in general there are three different interaction strengths between the ``up" and ``down" 
components, $g_{\uparrow\uparrow}$, $g_{\downarrow\downarrow}$, and $g_{\uparrow\downarrow}$, in 
the present problem we set $g_{\uparrow\uparrow} = g_{\downarrow\downarrow} = g$ and tune the 
ratio $g/g_{\uparrow \downarrow}$ (both assumed to be positive). Our total Hamiltonian is thus 
$H = H_0 + V_{\rm int}$, which yields two coupled Gross-Pitaevskii-like equations for the two 
components of the order parameter,
\begin{eqnarray}
\left[ \begin{array}{cc}
H_{11}& H_{12} \\
H_{12}^*& H_{22}  
\end{array} \right] 
\left[ \begin{array}{c}
\Phi_{\uparrow} \\ \Phi_{\downarrow}
\end{array} \right] = 
\mu \left[ \begin{array}{c}
 \Phi_{\uparrow} \\ \Phi_{\downarrow}
\end{array} \right],
\label{2gp}
\end{eqnarray}
where $H_{11} = \hbar^2 k_0^2/(2M) + p^2 /(2M) + V(\rho) + \hbar \Omega/2 + g |\Phi_{\uparrow}|^2 
+ g_{\uparrow \downarrow} |\Phi_{\downarrow}|^2$, $H_{22} = \hbar^2 k_0^2/(2M) + p^2 / (2M) + 
V(\rho) - \hbar \Omega/2 + g |\Phi_{\downarrow}|^2 + g_{\uparrow \downarrow} |\Phi_{\uparrow}|^2$, 
$H_{12} = -i \hbar k_0 p_x / M - i \hbar \delta/2$, and $\mu$ is the chemical potential, which is 
determined from the condition that the total occupancy of the two pseudospin components 
$N_{\uparrow}$ and $N_{\downarrow}$ is equal to the total number of atoms $N$.

For the specific form of the spin-orbit coupling considered and in the absence of any trapping 
potential and of interparticle interactions, the density distribution of the gas may be either 
striped, or homogeneous \cite{ho, li, lu, mar, limar}. This is determined by the ratio between
$\hbar \Omega$ and the recoil energy $E_R = \hbar^2 k_0^2/(2 M)$ \cite{lin}. If the ratio $\hbar 
\Omega/(4 E_R)$ is smaller or larger than unity, the system is in the striped, or in the 
homogeneous phase, respectively. Furthermore, the expectation value of $V_{SO}$ in some state 
$[\Phi_{\uparrow}, \Phi_{\downarrow}]^T$ is $E_{SO} - \hbar^2 k_0^2/(2 M) = \hbar \Omega/2 
(N_{\uparrow} - N_{\downarrow}) - (\hbar^2 k_0/M) \int \left(\Phi_{\uparrow}^* \, {\partial 
\Phi_{\downarrow}}/{\partial x} - \Phi_{\downarrow}^* \, {\partial \Phi_{\uparrow}}/{\partial x} 
\right) \, dx dy$. The term proportional to $k_{0}^2$ is constant. The term proportional to 
$\Omega$ is spatially-independent and is responsible for the population imbalance between the 
two components. The most ``interesting" is the spatially-dependent term, which is proportional 
to $k_0$. It is this term that gives rise to the striped phase. 

\section{Non-interacting problem in the limit of quasi-one-dimensional motion}

For very strong transverse confinement, the atoms reside in the ground state associated with the 
transverse degrees of freedom for motion along the annulus. This fact allows us to make an ansatz 
for $\Phi_{\uparrow/\downarrow}$, 
\begin{eqnarray}
\Phi_{\uparrow/\downarrow}(\rho, \theta) = \phi_0(\rho) \Psi_{\uparrow/\downarrow}(\theta), 
\label{ans}
\end{eqnarray}
with a Gaussian density distribution in the transverse direction, i.e., $n_0(\rho) = |\phi_0(\rho)|^2
= \exp[-(\rho - R)^2/a_0^2]/(\sqrt{\pi} a_0 R)$, where $a_0$ is the oscillator length, $a_0 = (\hbar/M 
\omega)^{1/2}$. Integrating over the transverse degrees of freedom and assuming that $R \gg a_0$ 
we thus develop the following quasi-one-dimensional eigenvalue problem for $\Psi_{\uparrow}$ and 
$\Psi_{\downarrow}$ (i.e., two coupled differential equations),
\begin{eqnarray}
\left( - \frac {\hbar^2} {2MR^2} \frac {\partial^2} {\partial \theta^2} 
+ \frac 1 2 \hbar \Omega - E \right) \Psi_{\uparrow}  +
\nonumber \\
 + \frac {\hbar^2 k_0} {M R} 
 \left( \frac 1 2 \cos \theta + \sin \theta \frac {\partial} {\partial \theta} \right) 
 \Psi_{\downarrow} = 0,
 \nonumber \\
 \left( - \frac {\hbar^2} {2MR^2} \frac {\partial^2} {\partial \theta^2} 
- \frac 1 2 \hbar \Omega - E \right) \Psi_{\downarrow}  +
\nonumber \\
 - \frac {\hbar^2 k_0} {M R} 
 \left( \frac 1 2 \cos \theta + \sin \theta \frac {\partial} {\partial \theta} \right) 
 \Psi_{\uparrow} = 0,
\label{eig123}
\end{eqnarray}
where $E$ is the eigenenergy.

In the absence of interactions there are three energy scales, $\hbar^2/(M R^2)$, $\hbar^2 k_0/
(M R)$, and $\hbar \Omega$. The above coupled equations have an analytic solution in the limit 
where $\hbar^2 k_0/(M R)$ is much smaller either than $\hbar^2/(M R^2)$, or than $\hbar \Omega$
(we stress that for $k_0 = 0$ the two equations decouple, while for ``small" values of 
$k_0 R$ they couple perturbatively). Defining the two dimensionless parameters $k_0 R \ll 1$ 
(i.e., the ratio between the second scale and the first), and $\lambda \equiv \hbar k_0/(M \Omega 
R)$ (i.e., the ratio between the second scale and the third), we find that for $k_0 R \ll 1$,
\begin{eqnarray}
\Psi_{\uparrow} &=& - \frac {k_0 R} {\sqrt{2 \pi}} \frac 1 {1 + {\epsilon}} \cos \theta,
\nonumber \\
\Psi_{\downarrow} &=& \frac {1} {\sqrt{2 \pi}} 
 \left[ 1 - \frac 3 8 (k_0 R)^2 (1 + {\epsilon}) \cos 2 \theta \right].
\label{an1}
\end{eqnarray}
The eigenenergy is $E/(\hbar \Omega) = -1/2 - (k_0 R)^2/[2 {\epsilon} (1 + \epsilon)]$,
where ${\epsilon} \equiv 2 M R^2 \Omega/\hbar$ is defined as the ratio between $\hbar \Omega$ 
and $\hbar^2/(2 M R^2)$. If $\alpha \equiv \hbar \Omega/E_R$, then $\epsilon = \alpha (k_0 R)^2$. 
For typical values of $\alpha$ of order unity, $\epsilon \sim (k_0 R)^2$. A second analytic 
solution is found for $\lambda \ll 1$,
\begin{eqnarray}
  \Psi_{\uparrow} &=& - \frac {\lambda} {\sqrt{8 \pi}} \cos \theta,
  \nonumber \\
  \Psi_{\downarrow} &=& \frac 1 {\sqrt{2 \pi}}
  \left[ 1 - \frac 3 {16} \lambda (k_0 R) \cos 2 \theta \right],
\label{an2}
\end{eqnarray}
while the eigenenergy is $E/(\hbar \Omega) = - 1/2 - \lambda^2/8$. 

Expanding the eigenfunctions in the basis of the plane-wave states, we have also solved the
eigenvalue problem numerically. The eigenfunctions of lowest energy have the general form 
$\Psi_{\uparrow} = \sum_m c_{2m} \cos 2 m \theta$ and $\Psi_{\downarrow} = \sum_m d_{2m+1} 
\cos(2 m + 1) \theta$, or vice versa, i.e., with $\Psi_{\uparrow}$ exchanged with 
$\Psi_{\downarrow}$. From these expressions some generic features follow for $\Psi_{\uparrow}$ 
and $\Psi_{\downarrow}$, which are also valid for the solutions found in Eqs.\,(\ref{an1}) and 
(\ref{an2}): (i) These solutions are even (the Hamiltonian is invariant under the transformation 
$\theta \to - \theta$ and thus the eigenstates are parity eigenstates). (ii) The density 
$n_{\downarrow}$ has nodes at $\theta = \pi/2$ and $3 \pi/2$, while at the same points 
$n_{\uparrow}$ has maxima, or vice versa. (iii) The periodicity of $\Psi_{\uparrow}$ is $\pi$ 
and of $\Psi_{\downarrow}$ it is $2 \pi$, or vice versa. (iv) Finally, the density of both 
components has a periodicity of $\pi$. 

Turning to the degeneracy of these solutions, the lowest-energy eigenstates are even. For all 
current experiments, the laser beam has a wavelength of order 1 $\mu$m, which means that 
$k_0 \sim 2 \pi$ ($\mu$m)$^{-1}$. For a typical value of $R \sim 10$ $\mu$m, $k_0 R$ is thus at
least 10, or larger. In the typical limiting case $k_0 R \gg 1$, the system is in the 
striped-like phase, the density variations are located around opposite ends of the ring, while 
the density effectively vanishes elsewhere [see, e.g., Fig. 1(e)]. There is thus a corresponding 
antisymmetric solution which becomes nearly degenerate with the symmetric one in this limit (with 
a corresponding two-fold degeneracy). In addition, the transformation $\Psi_{\uparrow} \to 
\Psi_{\downarrow}$, $\Psi_{\downarrow} \to -\Psi_{\uparrow}$ and $\Omega \to -\Omega$ leaves the 
eigenvalue problem unaffected. This gives rise to another two-fold (near) degeneracy for ``small" 
$\Omega$, which is exact for $\Omega = 0$. 
 
\begin{figure}
\includegraphics[width=5cm,height=4cm,angle=-0]{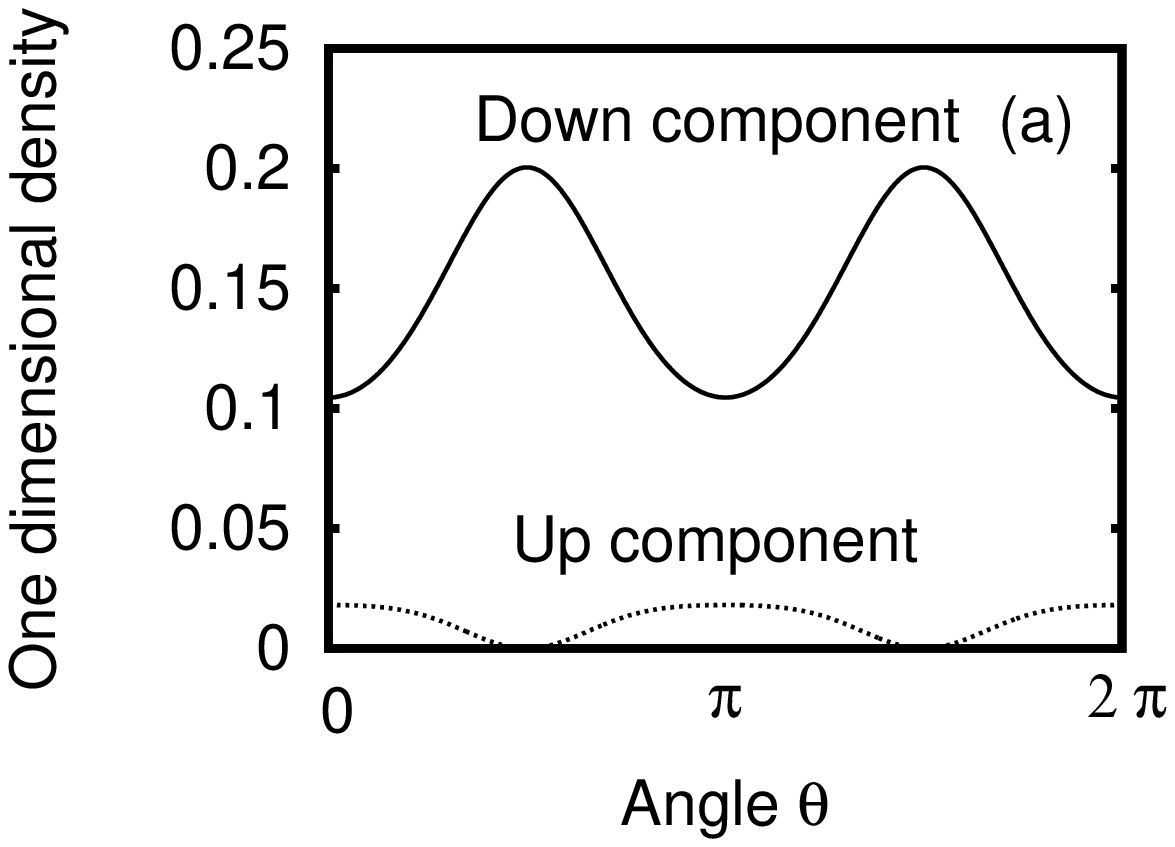}
\includegraphics[width=5cm,height=4cm,angle=-0]{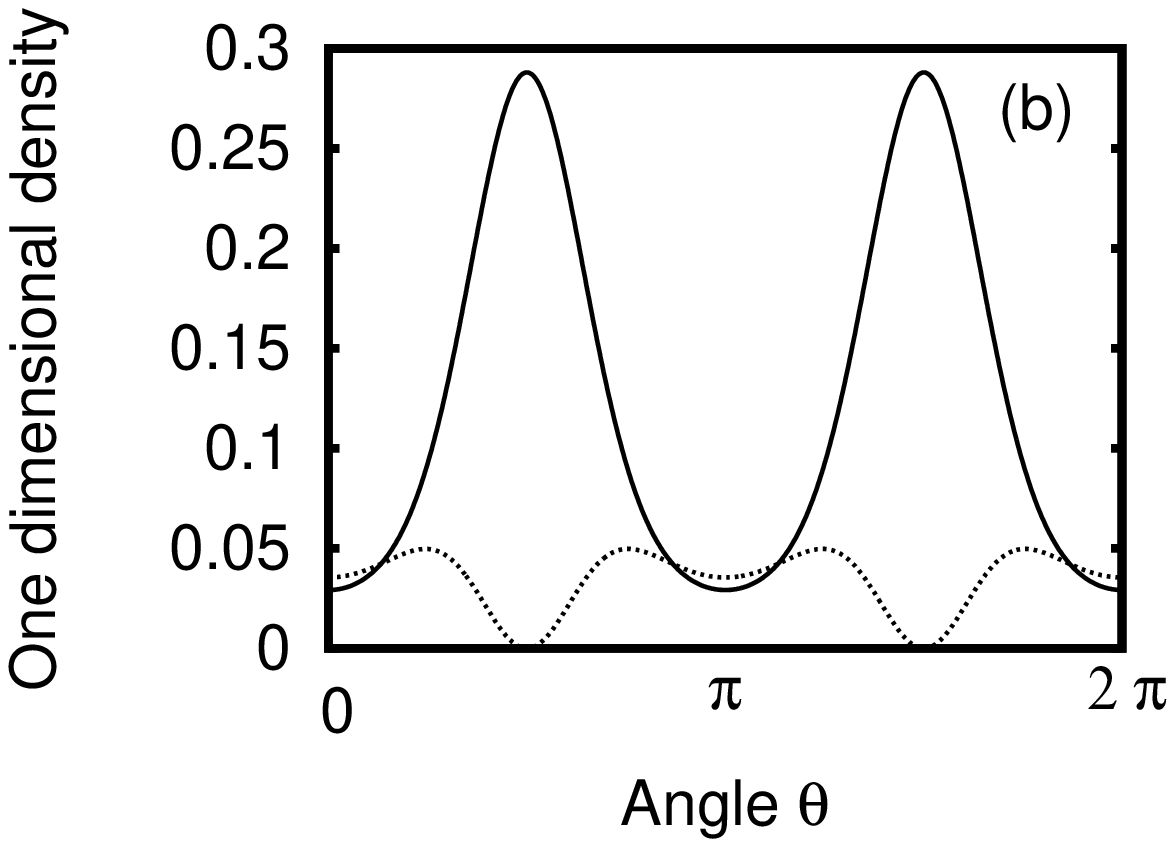}
\includegraphics[width=5cm,height=4cm,angle=-0]{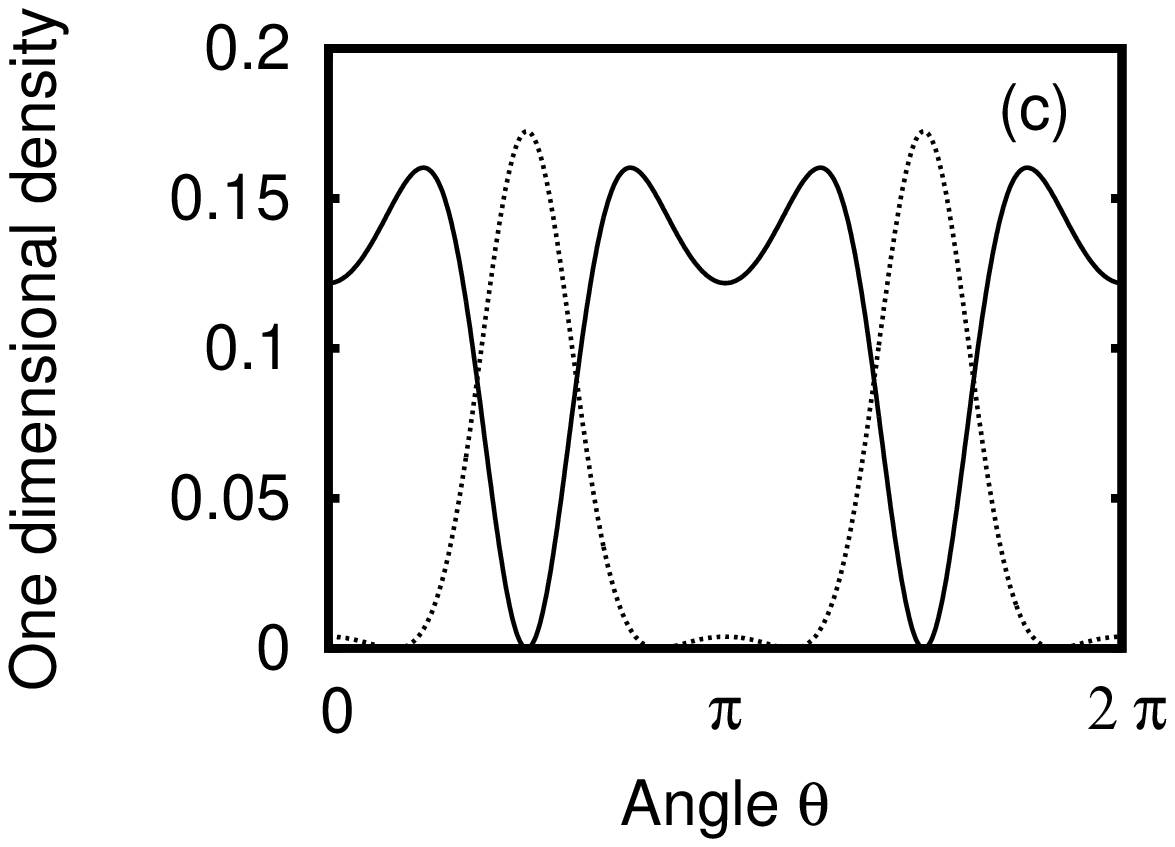}
\includegraphics[width=5cm,height=4cm,angle=-0]{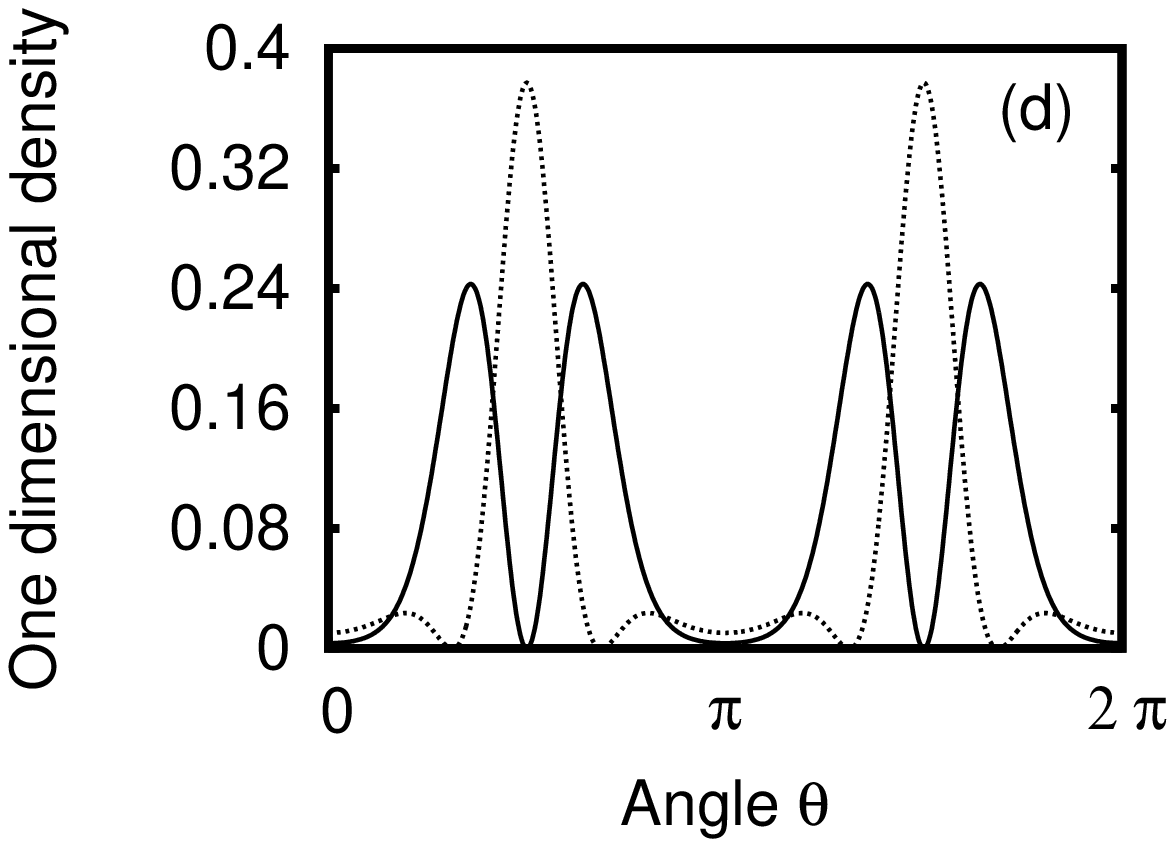}
\includegraphics[width=5cm,height=4cm,angle=-0]{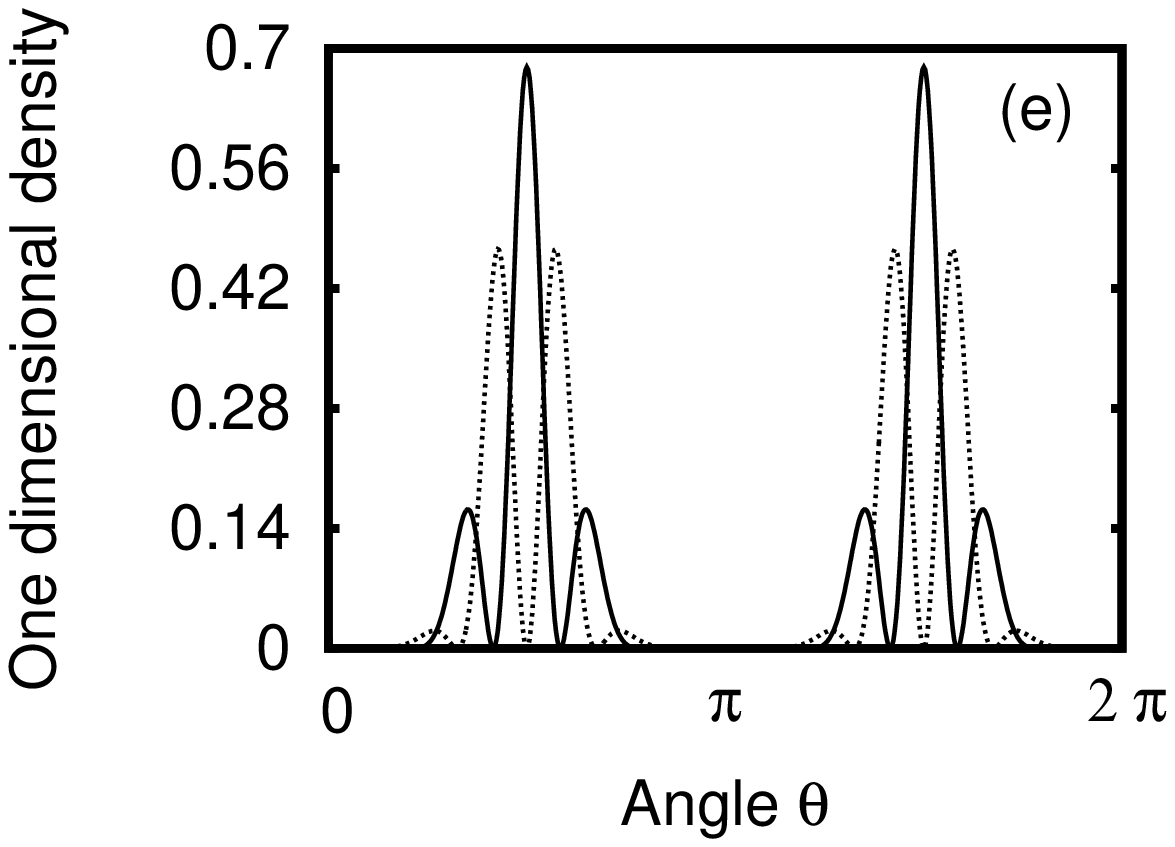}
\caption{One-dimensional density of the two pseudospin components, ``up" (dotted, lighter 
curve) and ``down" (solid, darker curve), within the quasi-one-dimensional model, for fixed 
values of $R$ and $\Omega$, with $\hbar/(M \Omega R^2) = 1$, and for an increasing value of $k_0$, 
$k_0 R = 1.0$ (a), 1.5 (b), 1.6 (c), 2.8 (d), and 6 (e).}
\end{figure}

The physical picture that emerges from the above calculation is that for $k_0 R \ll 1$ (and any 
value of $\Omega$), i.e., for sufficiently small values of $k_0$ and/or of $R$, as well as for 
$\hbar k_0/(M \Omega R) \ll 1$, i.e., for sufficiently small $k_0$ and/or large values of $\Omega$
and/or large values of $R$, the density variation of both components is sinusoidal. As $k_0 R$ 
increases (with all the other parameters fixed), the two components start to localize around 
$\theta = \pi/2$ and $\theta = 3 \pi/2$, with a pronounced density variation around these two 
points, which is the analogue of the striped phase. This transition is seen in Fig.\,1, where 
we plot the one-dimensional density of the two components for an increasing value of $k_0$, 
$k_0 R = 1.0$ (a), 1.5 (b) and 1.6 (c), 2.8 (d), and 6 (e), and for a fixed value of $\hbar/(M 
\Omega R^2) = 1$. The population imbalance also decreases with increasing $k_0$, since the term 
in the Hamiltonian that is proportional to $\Omega$ becomes smaller compared to the one that
is proportional to $k_0$. It is also remarkable that an increase of the value of $k_0 R$ 
of less than a factor of two results in a rather dramatic change in the density distribution of 
the two components, as seen, e.g., between Figs.\,1 (b) and 1 (c). As we show also below, 
this indicates a very rich structure, which is a generic feature of this problem.

In the first panel (a) of Fig.\,2 we have plotted the density $n_{\uparrow,\downarrow} \equiv 
|\Psi_{\uparrow,\downarrow}(\theta)|^2 \exp[-(\rho - R)^2/a_0^2]/(\sqrt{\pi} a_0 R)$, where 
$\Psi_{\uparrow, \downarrow}$ are the eigensolutions of Eqs.\,(\ref{eig123}). The parameters we 
have chosen are the ones which correspond to the results shown in the lower panels of Fig.\,2, 
namely $\hbar/(M \Omega R^2) = 1/16$, $\Omega/\omega = 1$, and $k_0 R = 0.4$ (first column), 
$\hbar/(M \Omega R^2) = 1/16$, $\Omega/\omega = 1$, and $k_0 R = 10$ (second column), and finally 
$\hbar/(M \Omega R^2) = 1/160$, $\Omega/\omega = 10$, and $k_0 R = 10$ (third column). Panel (a) 
of Fig.\,2 also serves as a reference to the lower panels which show solutions of the 
two-dimensional problem, as we discuss in the following section.

\section{Eigenvalue problem in the case of an annular potential}

We now investigate the effect of the finiteness of the width of the annulus, still in the absence 
of interactions. In this case, the width of the annulus (the oscillator length $a_0$) provides an 
extra length scale, which is a fraction of $R$. For a narrow annulus ($R \gg a_0$) one goes back 
to the problem of quasi-one-dimensional motion discussed above. When $a_0 \approx R$ then the 
problem reduces to that of a harmonic trapping potential \cite{xu, zha, har2, zha2, ram, aft, 
sinha, huram, huliu} with a small ``hole" in the center of the trap. The most interesting case is 
thus the intermediate one when $a_0 \lesssim R$.

We show in panel (b) of Fig.\,2 the two-dimensional density distribution of the two components 
that we have obtained numerically using the method of imaginary-time propagation (for the 
details of this calculation see the Appendix), considering an annular potential with $R/a_0 = 4$ 
and some representative values of $k_0$ and $\Omega$. In the first column, of ``small" $k_0$, 
$k_0 R = 0.4$, there is a large population imbalance between the two components  -- here $\hbar 
\Omega/E_R = 200$. While the ``up" component has a pronounced axial asymmetry, the dominant ``down" 
component also lacks axial symmetry with the maximum of the density being at $\theta = 0$ and $\pi$ 
(which is very weak, and is hardly visible in the plot). In the second column with a ``large" value 
of $k_0$, $k_0 R = 10$, both components show a striped phase and have an almost equal population; 
here $\hbar \Omega/E_R = 8/25$. Finally the phase shown in the third column with $\Omega/\omega = 
10$ and $k_0 R = 10$ resembles in a sense the phase with a sinusoidal density distribution found 
within the quasi-one-dimensional model, with a large population imbalance (here $\hbar \Omega/E_R 
= 16/5$). 

\begin{figure}
\includegraphics[width=8cm,height=16cm,angle=-0]{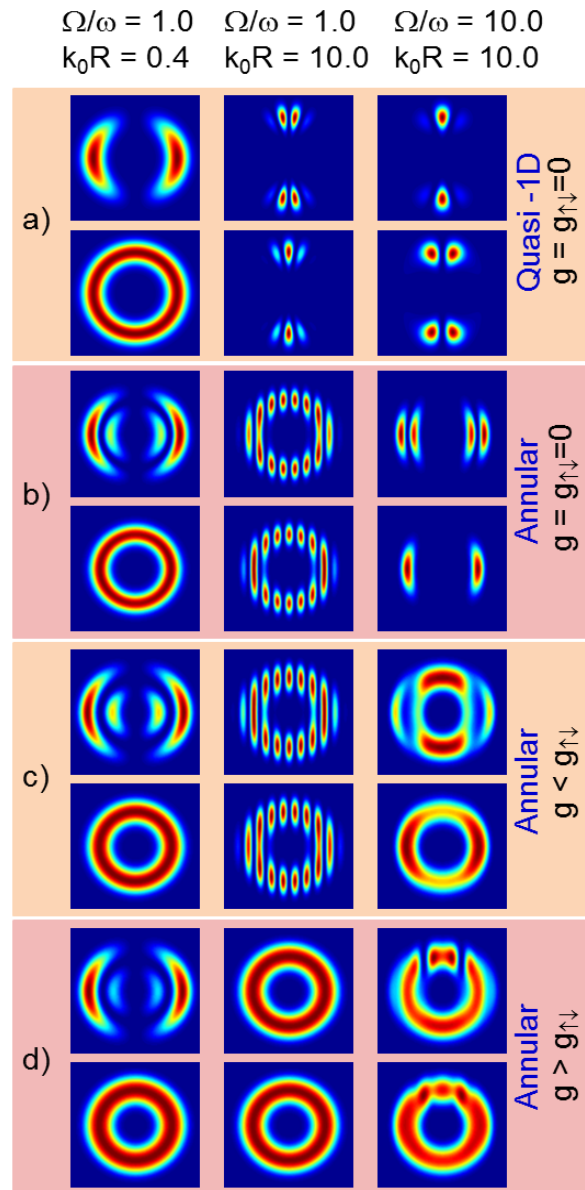}
\caption{Two-dimensional density of the two pseudospin components, ``up" (higher) and ``down" 
(lower). Panel (a) shows the result of the quasi-one-dimensional model with a Gaussian transverse 
profile. Panel (b) shows the density in an annular potential and in the absence of interactions. 
Panel (c) shows the result in an annular potential in the presence of interactions for $g < 
g_{\uparrow \downarrow}$ and panel (d) for $g > g_{\uparrow \downarrow}$. The color scale is not 
the same in all the plots, since in some of them there is a large population imbalance.}
\end{figure} 

\begin{figure}
\includegraphics[width=6.5cm,height=3cm,angle=-0]{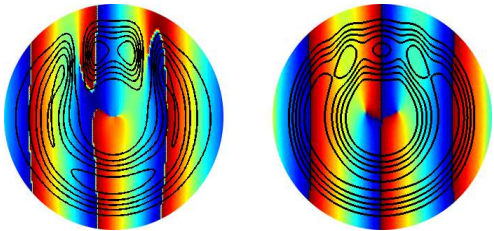}
\caption{Two-dimensional density (black contour lines) and phase of the two pseudospin components, 
``up" (left) and ``down" (right) for the data of the third column of Fig.\,2 (d). The color scale 
shows the phase with blue as zero and red as $2 \pi$.}
\end{figure} 

\section{Effect of the interactions}

Having established a rather complete picture about the single-particle eigenvalue problem, we 
now turn to the effect of the interactions. In a homogeneous system when $g_{\uparrow \downarrow} > 
g$ the striped phase (where also the density minima/maxima of the one component coincide with the 
maxima/minima of the other) is energetically more favourable, since this configuration minimizes the 
overlap between the two components [due to the last term in Eq.\,(\ref{avb}) below]. This may be seen 
if one writes the interaction energy in the following form 
\begin{eqnarray}
E_{\rm int} = (g/2) \int n^2({\bf r}) \, d{\bf r} + (g_{\uparrow 
\downarrow} - g) \int n_{\uparrow}({\bf r}) n_{\downarrow}({\bf r}) \, d{\bf r}, 
\label{avb}
\end{eqnarray}
where $n({\bf r}) = n_{\uparrow}({\bf r}) + n_{\downarrow}({\bf r})$ is the total density. According 
to the numerical results that we have obtained using the method of imaginary-time propagation, when 
$g_{\uparrow \downarrow} > g$ the interactions ``combine" with the (spatially dependent) term of the 
spin-orbit Hamiltonian that is proportional to $k_0$ (and favours the formation of stripes, as we 
argued earlier). Both terms make it energetically favourable for the system to form a striped-like 
phase (similar to the one that we described in the case of an annular potential with spin-orbit 
coupling, but without interactions). Panel (c) of Fig.\,2 shows $n_{\uparrow}$ and $n_{\downarrow}$, 
for $g/g_{\uparrow \downarrow}=1/2$, and $Ng/(2 \pi R a_0 \hbar \omega) = 25/(4 \pi)$, or $Ng/
(2 \pi R a_0 E_R) = 1250/\pi$ in the first column and $Ng/(2 \pi R a_0 E_R) = 2/\pi$ in the second 
and in the third column. While in the first two columns the density is slightly affected by the 
interactions [i.e., as compared to the density in Fig.\,2 (b)], in the third column the interactions 
expand the two components more around the annulus.

In panel (d) of Fig.\,2 we have chosen  $g/g_{\uparrow \downarrow}=2$. The first column indicates 
that the density is not affected drastically by the interactions. On the other hand, in the second
column we see that the density is essentially axially symmetric, in contrast to the striped-like phase, 
which is seen in the absence of interactions in the second column of panel (b). Thus, this phase is 
strongly affected by the interactions. 

At this point it is interesting to make contact between the results of Fig.\,2 and those of 
no trapping potential. As we mentioned also above, in the absence of interactions when the ratio 
$\hbar\Omega/(4 E_R)$ is smaller or larger than unity, the system is in the striped, or in the 
homogeneous phase, respectively \cite{lin}. The actual value of this ratio is 50 (first column), 
0.08 (second column), and 0.8 (third column). According to the above criterion, the first and the 
second columns of Fig.\,2 (b) are in the homogeneous and in the striped phases, respectively, which 
is consistent with our data. For the interacting problem, the critical value for the ratio $\hbar 
\Omega/(4 E_R)$ is $0.19/4 = 0.0475$ in this case \cite{lin, li, lu, mar, limar}. Therefore, the 
first and the third columns of Fig.\,2(c) and (d) are clearly in the homogeneous phase, which again 
is consistent with our data.

In the third column of Fig.\,2 (d) we show a solution of nonzero circulation, which is only 
possible in the presence of interactions and for a wide enough annulus. For the same data, Fig.\,3 
shows the phase of the two order parameters along with the corresponding density. As seen in this 
plot, there are two phase singularities in the ``down" component and three in the ``up". In the 
``down" component the corresponding vortices reside in the central region of exponentially small 
density, while in the ``up" component one of the three is in the central region and two in the
region of nonzero density (for this reason the deviation of the density of the ``up" component 
from axial symmetry is more pronounced than that of the ``down" component). For the same reason, 
while all the other plots of Fig.\,2 have a mirror symmetry with respect to the $x$ axis, this 
symmetry is broken. We should also mention that the sense of circulation of the state of the third 
column of Fig.\,2 (d) is determined by the initial condition that is given in the (iterative) method 
of imaginary-time propagation. Furthermore, with a ``mean" density $n_0 = N/(2 \pi R a_0)$, for the 
healing length $\xi_0$ that corresponds to the coupling constant $g$, $\xi_0/a_0 = \sqrt{\hbar \omega/
(2 n_0 g)} = \sqrt{2 \pi/25} \approx 1/2$. Therefore, $\xi_0$ is roughly one half of the width of the 
annulus, since $a_0$ sets this scale. Clearly these length scales play a crucial role, since the 
vortex states we have found have to ``fit" within the annulus and this becomes possible provided that 
the width of the annulus is wide enough, i.e., it is comparable, or larger than the coherence length 
$\xi_0$. 

The vortex states we have found in Fig.\,3 involve two large parameters: Rabi coupling $\Omega/\omega 
= 10$ and Raman wavenumber $k_0 R = 10$, as well and the important effect of interactions with $g > 
g_{\uparrow\downarrow}$. Each of these plays an important role, since the vortex states appear only in 
the right column of Fig.\,2(d).  As seen in Fig.\,3, there are persistent currents in both the upper
component (arising from the phase singularity in the central region) and the lower component (arising 
from the two phase singularities in the central region). In addition, the upper component has two 
vortices in the annular region with finite number density. Such states \cite{sinha, huram, ramopa, 
fetter} have the potential to serve as cold-atom analogs of ring currents and definitely merit further 
more detailed investigation.


\section{Conclusions}

Spin-orbit coupled Bose-Einstein-condensed atoms confined in an annular potential show a 
variety of phases as a result of the (single-particle effect of the) spin-orbit coupling 
and of the (many-body effect of the) coupling between the two atoms, while the non-trivial 
topology of the annular potential is another novel aspect of the problem that we have solved. 

Figure 1 summarizes the results for the solutions of the eigenvalue problem in the case of a 
very narrow annulus, as $k_0$ increases (from top to bottom). The homogeneous phase that is 
present in an untrapped system is replaced by a sinusoidal density variation in the case of 
a very narrow annulus, which evolves continuously into a striped-like phase as $k_0$ increases. 
As the width of the annulus increases, this picture persists qualitatively, as seen in Fig.\,2, 
where interactions have also been included. Depending on the relative strength of $g$ and 
$g_{\uparrow \downarrow}$ the interactions either favor the striped phase, or suppress it, 
while they may also give rise to nonlinear solutions with a nonzero circulation that require 
nonzero interactions. The very interesting implication of these solutions is that in a spin-orbit 
coupled system vortex states might spontaneously be created in one or both components, sometimes 
in the physical region with nonzero density and sometimes in the central region with zero density 
(these latter vortices lead to quantized circulation around the annulus).

While our study does not in any way exhaust all the possible phases, it gives a sense of the 
richness of this problem. A natural future project is a more systematic study of the solutions 
presented in Fig.\,2 (d) and Fig.\,3. Finally, the dimensional reduction we have performed 
provides a general method, which (with the trivial inclusion of interactions) may be useful in 
investigating questions related with e.g., nonlinear, solitary-wave solutions in quasi-one-dimensional 
spin-orbit coupled systems. 

\acknowledgements

E. \"O. K., G. M. K., and F. M. acknowledge the hospitality of Lund University. This work was partially 
financed by the Swedish Research Council and ``NanoLund".

\appendix*

\section{Numerical method}

The numerical method used to solve the two coupled Gross-Pitaevskii-like equations for the two pseudospin 
components of the order parameter, given by Eq. (\ref{2gp}) of the main text, is based on a fourth-order 
split-step Fourier method within an imaginary-time propagation technique. The details of the method were 
previously discussed in Ref.\,\cite{chin05}. Therefore here we give only a brief overview of the method. 

The starting point, which forms the basis of the imaginary-time propagation method, is to consider the 
time-dependent Schr\"odinger equation in imaginary time, i.e., $\tau=-it$ (in the expressions below we 
use atomic units for convenience)
\begin{equation} \label{eq:schrode}
\frac{\partial \psi(\tau)}{\partial \tau} = H \psi(\tau),
\end{equation}
where $H$ is the Hamiltonian, and to propagate the wave function $\psi$ under the action of a time evolution 
operator $\exp(-\tau H)$,
\begin{eqnarray} \label{eq:prop}
\psi(\tau) \propto \exp(-\tau H)\psi(0).
\end{eqnarray}
If the initial state $\psi(0)$ is expanded in the eigenstates $\phi_i$ of the Hamiltonian, with eigenvalues 
$\varepsilon_i$, then 
\begin{equation} \label{eq:expansion}
\psi(\tau) \propto \sum_{i} c_i\phi_i\exp(-\tau\varepsilon_i).
\end{equation}
Therefore, the time-evolution operator leads to an exponential decay of the eigenstates, where the one with 
the lowest energy $\varepsilon_0$ decays with the slowest rate and $\psi(\tau)$ eventually converges to the 
ground state $\phi_0$ as $\tau \to \infty$.

To evaluate the time evolution given in Eq.\,(\ref{eq:expansion}), we use a forward fourth-order algorithm 
called split-step Fourier method: 
\begin{eqnarray} \label{eq:fourthorder}
\psi(\Delta\tau) &=& e^{-(1/6)\Delta\tau V(\Delta\tau)}e^{-(1/2)\Delta\tau T}e^{-(2/3)\Delta\tau 
\tilde{V}(\Delta\tau/2)} 
\times \nonumber \\
&\times& e^{-(1/2)\Delta\tau T}e^{-(1/6)\Delta\tau V(0)}\psi(0).
\end{eqnarray}
In our study $V = H - K$, where $K$ is the kinetic energy, serves as an effective potential and it consists of the 
external trapping potential, the interaction and the spin-orbit coupling terms. The mid-point effective potential 
in Eq.\,(\ref{eq:fourthorder}) is given by 
\begin{equation} \label{eq:mid-point}
\tilde{V}=V+\frac{\Delta\tau^2}{48}[V,[T,V]].
\end{equation}
The order parameter is discretized on a square mesh of $2^{\cal N}$ points (${\cal N}$ is an integer) in both $x$ 
and $y$ directions with separation $\Delta l$ and its Fourier transform is computed using the discretized fast 
Fourier transform (FFT). 

In our calculations we have used $256^2$ grid points with $\Delta l=0.1$, where we have observed that larger grid 
sizes do not have a significant effect on the obtained results. The time step $\Delta\tau$ had to be chosen small 
enough to ensure that the Hamiltonian of the system does not change dramatically between the iterations. The optimum 
value of $\Delta\tau$, which satisfies this condition and also provides a good convergence in our study, has been 
determined as 0.01. On the other hand, we have observed that the use of smaller time steps does not improve the 
accuracy of our results. 

The time evolution procedure requires an initial condition for the wave function that is propagated in imaginary 
time until a steady-state solution with the (local or absolute) minimum of energy is reached after a sufficiently 
large number of iterations. For example, to obtain the solution with nonzero circulation given in the third column 
of Fig.\, 2 (d) and in Fig.\,3, we have performed calculations using ten different initial states. We have 
found that all the results exhibit density distributions with a nonzero circulation and the result presented in 
Fig.\,2 (d) and in Fig.\,3 is the energetically most favorable state.


\begin{thebibliography}{99}

\bibitem{dre} G. Dresselhaus, Phys. Rev. {\bf 100}, 580 (1955); Y. A. Bychkov and E. I. Rashba, 
J. Phys. Chem. {\bf 17}, 6039 (1984).

\bibitem{mir} F. Mireles and G. Kirczenow, Phys. Rev. B {\bf 64}, 024426 (2001); L. W. Molenkamp, 
G. Schmidt, and G. E. W. Bauer, {\it ibid.} {\bf 64}, 121202 (2001).

\bibitem{kas} A. Kasic, M. Schubert, S. Einfeldt, D. Hommel, and T. E. Tiwald, Phys. Rev. B 
{\bf 62}, 7365 (2000).

\bibitem{and} T. Ando, J. Phys. Soc. Jpn. {\bf 69}, 1757 (2000); C. L. Kane and E. J. Mele, Phys. 
Rev. Lett. {\bf 95}, 226801 (2005); H. Min, J. E. Hill, N. A. Sinitsyn, B. R. Sahu, L. Kleinman, 
and A. H. MacDonald, Phys. Rev. B {\bf 74}, 165310 (2006).

\bibitem{gov} M. Governale, Phy. Rev. Lett. {\bf 89}, 20682 (2002).

\bibitem{wei} S. Weiss and R. Egger, Phy. Rev. B {\bf 72}, 245301 (2005).

\bibitem{lip} E. Lipparini, M. Barranco, F. Malet, and M. Pi, Phy. Rev. B {\bf 79}, 115310 (2009).

\bibitem{nuclei} D. R. Inglis, Phys. Rev. {\bf 50}, 783 (1936); D. R. Inglis, Phys. Rev. 
{\bf 75}, 1767 (1949); L. S. Kisslinger, Phys. Rev. {\bf 104}, 1077 (1956).

\bibitem{zut} I. Zutic, J. Fabian, and S. Das Sarma, Rev. Mod. Phys. {\bf 76}, 323 (2004); X. L. 
Qi and S. C. Zhang, Physics Today, {\bf 63}, 33 (2010); M. Z. Hasan and C. L. Kane, Rev. Mod. Phys. 
{\bf 82}, 3045 (2010); X.-L. Qi and S.-C. Zhang, {\it ibid.} {\bf 83}, 1057 (2011).

\bibitem{at1} B. Seaman, M. Kr\"{a}mer, D. Anderson, and M. Holland, Phys. Rev. A {\bf 75}, 023615
(2007).

\bibitem{at2} R. Pepino, J. Cooper, D. Anderson, and M. Holland, Phys. Rev. Lett. {\bf 103}, 140405
(2009).

\bibitem{at3} P. Schlagheck, F. Malet, J. C. Cremon, and S. M. Reimann, New J. Phys. {\bf 12}, 065020
(2010).

\bibitem{at4} R. Pepino, J. Cooper, D. Meiser, D. Anderson, and M. Holland, Phys. Rev. A {\bf 82}, 
013640 (2010).

\bibitem{at5} A. Ramanathan, K. C. Wright, S. R. Muniz, M. Zelan, W. T. Hill, C. J. Lobb, K. Helmerson,
W. D. Phillips, and G. K. Campbell, Phys. Rev. Lett. {\bf 106}, 130401 (2011).

\bibitem{at6} J. Brantut, J. Meineke, D. Stadler, S. Krinner, and T. Esslinger, Science {\bf 337}, 1069 
(2012).

\bibitem{at7} L. H. Kristinsd\'ottir, O. Karlstr\"om, J. Bjerlin, J. C. Cremon, P. Schlagheck, A. Wacker, 
and S. M. Reimann, Phys. Rev. Lett. {\bf 110}, 085303 (2013).

\bibitem{at8} M. Edwards, Nature Phys. {\bf 9}, 68 (2013).

\bibitem{at9} M. K. Olsen and A. S. Bradley, Phys. Rev. A {\bf 91}, 043635 (2015).

\bibitem{spintr} D. D. Awschalom and M. E. Flatte, Nature Phys. {\bf 3}, 153 (2007); D. D. Awschalom and 
N. Samarth, Physics {\bf 2}, 50 (2009); D. D. Awschalom, Lee C. Bassett, A. S. Dzurak, E. L. Hu, and 
J. R. Petta, Science {\bf 339}, 1174 (2013).

\bibitem{ost} K. Osterloh, M. Baig, L. Santos, P. Zoller, and M. Lewenstein, Phys. Rev. Lett. 
{\bf 95}, 010403 (2005); J. Ruseckas, G. Juzeli$\bar{\rm{u}}$nas, P. \"Ohberg, and M. Fleischhauer, {\it ibid.} 
{\bf 95}, 010404 (2005); T. D. Stanescu, B. Anderson, and V. Galitski, Phy. Rev. A {\bf 78}, 
023616 (2008).

\bibitem{lin0} Y.-J. Lin, R. L. Compton, K. Jim\'enez-Garc\'ia, J. V. Porto, and I. B. Spielman, 
Nature (London) {\bf 462}, 628 (2009).

\bibitem{lin} Y.-J. Lin, K. Jim\'enez-Garc\'ia, and I. B. Spielman, Nature (London) {\bf 471}, 
83 (2011).

\bibitem{wang} P. Wang, Z.-Q. Yu, Z. Fu, J. Miao, L. Huang, S. Chai, H.Zhai, and J.Zhang, Phys. Rev. 
Lett. {\bf 109}, 095301 (2012); L. W. Cheuk, A. T. Sommer, Z. Hadzibabic, T. Yefsah, W. S. Bakr, and 
M. W. Zwierlein, Phys. Rev. Lett. {\bf 109}, 095302 (2012).

\bibitem{spth} I. B. Spielman, Phys. Rev. A {\bf 79}, 063613 (2009).

\bibitem{merkl} M. Merkl, G. Juzeli$\bar{\rm{u}}$nas, and P. \"Ohberg, Eur. Phys. J. D {\bf 59}, 257 (2010).

\bibitem{wan} C. Wang, C. Gao, C.-M. Jian, and H. Zhai, Phys. Rev. Lett., {\bf 105}, 160403 
(2010).

\bibitem{ho} T.-L. Ho and S. Zhang, Phys. Rev. Lett. {\bf 107}, 150403 (2011).

\bibitem{xu} X.-Q. Xu and J. Hoon Han, Phys. Rev. Lett. {\bf 107}, 200401 (2011).

\bibitem{mxu2} Z. F. Xu, R. L\"u, and L. You, Phys. Rev. A {\bf 83}, 053602 (2011); T. Kawakami, 
T. Mizushima, and K. Machida, {\it ibid.} {\bf 84}, 011607 (2011).

\bibitem{rad} J. Radic, T. A. Sedrakyan, I. B. Spielman, and V. Galitski, Phys. Rev. A {\bf 84}, 
063604 (2011).

\bibitem{wu11} C.-J. Wu, I. Mondragon-Shem, and X.–F. Zhou, Chinese Phys. Lett. {\bf 28}, 097102 (2011).

\bibitem{zhou11} X.-F. Zhou, J. Zhou, and C. Wu, Phys. Rev. A {\bf 84}, 063624 (2011).

\bibitem{zha} Y. Zhang, L. Mao, and C. Zhang, Phys. Rev. Lett. {\bf 108}, 035302 (2012).

\bibitem{li} Y. Li, L. P. Pitaevskii, and S. Stringari, Phys. Rev. Lett. {\bf 108}, 225301 (2012).

\bibitem{and2} B. M. Anderson, G. Juzeli$\bar{\rm{u}}$nas, V. M. Galitski, and I. B. Spielman, Phys. Rev. Lett. 
{\bf 108}, 235301 (2012).

\bibitem{har2} Wei Zheng and Zhibing Li, Phys. Rev. A {\bf 85}, 053607 (2012).

\bibitem{li11} Y. Li, X. Zhou, and C. Wu, e-print arXiv 1205.2162.

\bibitem{gal} V. Galitski and I. B. Spielman, Nature (London) {\bf 494}, 49 (2013).

\bibitem{zha2} Y. Zhang, G. Chen, and C. Zhang, Scientific Reports {\bf 3}, 1937 (2013).

\bibitem{ram} B. Ramachandhran, H. Hu, and H. Pu, Phys. Rev. A {\bf 87}, 033627 (2013).

\bibitem{and3} B. M. Anderson, I. B. Spielman, and G. Juzeli$\bar{\rm{u}}$nas, Phys. Rev. Lett. {\bf 111}, 
125301 (2013).

\bibitem{xu3} Z.-F. Xu, L. You, and M. Ueda, Phys. Rev. A {\bf 87}, 063634 (2013).

\bibitem{xu4} Z.-F. Xu, S. Kobayashi, and M. Ueda, Phys. Rev. A {\bf 88}, 013621 (2013).

\bibitem{aft} A. Aftalion and P. Mason, Phys. Rev. A {\bf 88}, 023610 (2013).

\bibitem{den} Y. Deng, J. Cheng, H. Jing, C.-P. Sun, and S. Yi, Phys. Rev. Lett. {\bf 108}, 
125301 (2012); R. M. Wilson, B. M. Anderson, and C. W. Clark, {\it ibid.} {\bf 111}, 185303 
(2013). 

\bibitem{zhou22} X. Zhou, Y. Li, Z. Cai, and C. Wu, J. Phys. B {\bf 46}, 134001 (2013).

\bibitem{zez} D. A. Zezyulin, R. Driben, V. V. Konotop, and B. A. Malomed Phys. Rev. A {\bf 88}, 
013607 (2013).

\bibitem{che} X. Chen, M. Rabinovic, B. M. Anderson, and L. Santos, Phys. Rev. A {\bf 90}, 043632 
(2014).

\bibitem{kar11} Y. V. Kartashov, V. V. Konotop, and D. A. Zezyulin, Europh. Lett. {\bf 107}, 50002 
(2014).

\bibitem{lar} J. Larson, J.-P. Martikainen, A. Collin, and E. Sj\"oqvist, Phys. Rev. A {\bf 82}, 
043620 (2010); Y. Cheng, G. Tang, and S. K. Adhikari, {\it ibid.} {\bf 89}, 063602 (2014); Y. Zhang, 
Y. Xu, and T. Busch, {\it ibid.} {\bf 91}, 043629 (2015).

\bibitem{qu} C. Qu, K. Sun, and C. Zhang, Phys. Rev. A {\bf 91}, 053630 (2015); K. Sun, C. Qu, 
and C. Zhang, {\it ibid.} {\bf 91}, 063627 (2015).

\bibitem{gup} S. Gupta, K. W. Murch, K. L. Moore, T. P. Purdy, and D. M. Stamper-Kurn, Phys. 
Rev. Lett. {\bf 95}, 143201 (2005).

\bibitem{murray} N. Murray, M. Krygier, M. Edwards, K. C. Wright, G. K. Campbell, and C. W. Clark, 
Phys. Rev. A {\bf 88}, 053615 (2013).

\bibitem{wright} K. C. Wright, R. B. Blakestad, C. J. Lobb, W. D. Phillips, and G. K. Campbell, 
Phys. Rev. A {\bf 88}, 063633 (2013).

\bibitem{jia} C.-M. Jian and H. Zhai, Phys. Rev. B {\bf 84}, 060508 (2011).

\bibitem{zhai} H. Zhai, Int. J. Mod. Phys. B {\bf 26}, 1230001 (2012).
	
\bibitem{oza} T. Ozawa and G. Baym, Phys. Rev. A {\bf 85}, 063623 (2012).

\bibitem{oza2} T. Ozawa and G. Baym, Phys. Rev. Lett. {\bf 110}, 085304 (2013).

\bibitem{sinha} S. Sinha, R. Nath and L. Santos, Phys. Rev. Lett. {\bf 107}, 270401 (2011).

\bibitem{huram} H. Hu, B. Ramachandhran, H. Pu, and X.-J. Liu, Phys. Rev. Lett. {\bf 108}, 
010402 (2012).

\bibitem{huliu} H. Hu and X.-J. Liu, Phys. Rev. A {\bf 85}, 013619 (2012).

\bibitem{lu} Q.-Q. L\"u and D. E. Sheehy, Phys. Rev. A {\bf 88}, 043645 (2013).

\bibitem{mar} G. I. Martone, Y. Li, and S. Stringari, Phys. Rev. A {\bf 90}, 041604 (2014).

\bibitem{limar} Y. Li, G. I. Martone, and S. Stringari, e-print arXiv: 1410.5526.

\bibitem{ramopa} B. Ramachandhran, B. Opanchuk, X.-J. Liu, H. Pu, P. D. Drummond, and H. Hu, 
Phys. Rev. A {\bf 85}, 023606 (2012).

\bibitem{fetter} A. L. Fetter, Phys. Rev. A {\bf 89}, 023629 (2014).

\bibitem{chin05} S. A. Chin, and E. Krotscheck, Phys. Rev. E {\bf 72}, 036705 (2005).

\end{thebibliography}
\end{document}